\documentclass[
               showpacs,            
               preprintnumbers,     
               aps,                 
               prd,                 
               a4paper,             
               superscriptaddress,  
               nofootinbib,         
               tightenlines,        
               floats,floatfix      
               ]{revtex4}
\usepackage[dvips]{graphicx}
\usepackage{bm}
\usepackage{latexsym}
\usepackage{
amssymb}        

\newcommand{\lsim}{\mbox{\raisebox{-.9ex}{~$\stackrel{\mbox{$<$}}{\sim}$~}}}
\newcommand{\gsim}{\mbox{\raisebox{-.9ex}{~$\stackrel{\mbox{$>$}}{\sim}$~}}}

\renewcommand\({\left(}
\renewcommand\){\right)}
\renewcommand\[{\left[}
\renewcommand\]{\right]}

\newcommand\eqref[1]{(\ref{#1})}

\newcommand\ee{\end{equation}}
\newcommand\be{\begin{equation}}
\newcommand\eea{\end{eqnarray}}
\newcommand\bea{\begin{eqnarray}}

\def\cala{{\cal A}}

\def\calh{{\cal H}}

\def\calr{{\cal R}}

\newcommand{\seno}{{\rm sin}}
\newcommand{\coseno}{{\rm cos}}

\begin{document}

\title{Boosted perturbations at the end of inflation}
\author{Ignacio Zaballa}
\affiliation{Korea Institute for Advanced Study, Seoul 130-722, Korea.}
\author{Misao Sasaki}
\affiliation{Yukawa Institute for Theoretical Physics, Kyoto University, Kyoto 
606-8502, Japan.}
\affiliation{Korea Institute for Advanced Study, Seoul 130-722, Korea.}
\date{\today}
\pacs{04.70.-s, 98.80.-k \hfill} 
\preprint{YITP-09-71}

\begin{abstract}
We study the effect on the primordial cosmological perturbations
of a sharp transition from inflationary to a radiation and matter
dominated epoch respectively.
We assume that the perturbations are generated 
by the vacuum fluctuations of a scalar field slowly rolling down
its potential, and that the transition into the subsequent epoch
takes place much faster than a Hubble time.
The behaviour of the superhorizon perturbations corresponding to 
cosmological scales in this case is well known. 
However, it is not clear how perturbations on scales of and smaller than 
the Hubble horizon scale at the end of inflation
 may evolve through such a transition. 
We derive the evolution equation for the gravitational
potential $\Psi$, which allows us to study  the evolution of the
perturbations on all scales under these circumstances. 
We show that for a certain range of scales inside the horizon
at the end of inflation, the amplitude of the perturbations are
enhanced relative to the superhorizon scales.
This enhancement may lead to the overproduction of
Primordial Black Holes (PBHs), and therefore constrain the
dynamics of the transitions that take place at the end of inflation. 
\end{abstract}

\maketitle

\section{Introduction}

Inflation has become the main paradigm to understand the
presence of small inhomogeneities during the early universe.
The observed Cosmic Microwave Background (CMB) and 
Large Scale Structure (LSS) spectrum of primordial
perturbations, reveals that simple inflationary models are
consistent with the analysis of the current data.
Nevertheless, CMB and LSS observations only probe 
primordial perturbations on scales that leave the horizon
long before the end of inflation, and possible cosmological 
implications of the perturbations on subhorizon scales
have not been studied much. 

In this paper, we investigate the behavior of curvature perturbations
on scales of and smaller than the horizon size at the end of inflation.
We assume standard slow-roll inflation and that the perturbations 
are generated from the vacuum fluctuations of the inflaton field.
It is expected that the universe enters either a stage of
radiation-domination or matter-domination right after inflation,
depending on the efficiency of preheating or reheating, and the transition
takes place within a timescale shorter than the Hubble time.
We therefore model this transition as a change of the equation
of state from $w\approx -1$ to $w=1/3$ or to $w=0$ within a
time scale $\Delta t\ll H_e^{-1}$ where $H_e$ is the Hubble parameter
at the end of inflation. 

On large scales where the wavelengths are much larger than $\Delta t$,
we may regard the transition as instantaneous, occuring on
a given spacelike hypersurface.
Then we can match the perturbations between inflation and 
the succeeding epoch by requiring that the intrinsic and extrinsic curvatures
are continuous on the hypersurface at which we do the
matching~\cite{Israel:1966rt,Deruelle:1995kd,Martin:1997zd}.
It is natural to assume that the end of inflation occurs when the inflaton 
field reaches a critical value, $\phi_c(x,\tau)={\rm constant}$.
This implies that the transition occurs on a comoving surface
on which $\phi$ is spatially homogeneous.
Then one can show that we have the following junction conditions,
\bea
\[\calr_{\bm k}\]^+_-&=&0
\label{junc1}\,,
\\
\[\Psi_{\bm k}\]^+_-&=&0
\label{junc2}\,,
\eea
where $\calr$ is the curvature perturbation on comoving 
hypersurfaces, and $\Psi$ is the generalized gravitational
potential.
The junction conditions give the initial conditions for the
following radiation- or matter-dominated stage.
On superhorizon scales, the potential $\Psi$ then reaches a constant value,
$\Psi\sim\calr$, a few Hubble times after the end of
inflation~\cite{Deruelle:1995kd,Martin:1997zd,Lyth:2005ze}.

What we are interested in here is perturbations on scales of or 
smaller than the horizon scale. On subhorizons scales, but still larger 
than the scale $\Delta t$, the resulting perturbations are  the same 
as in the case of superhorizon perturbations if the universe is 
matter-dominated after the transition.
On the other hand, the result is not so
simple if the universe is radiation-dominated after the transion.
In Refs.~\cite{Lyth:2005ze, Zaballa:2006kh}, we investigated this
case, and argued there that the relative large amplitude of the 
potential $\Psi$ can be achieved on subhorizon scales and 
may lead to the overproduction of PBHs at the end of inflation. 
The constraint on the amplitude of the curvature perturbation 
is then tighter than the current bound from the PBHs formed
throughout the radiation epoch \cite{Zaballa:2006kh}
(for the constraints on the "standard" formation of PBHs
see~\cite{Carr:1993,Carr:1994ar,Kim:1996hr,Green:1997sz,
Leach:2000ea,Bugaev:2006fe,Kohri:2007qn,Peiris:2008be,Josan:2009qn,
Bugaev:2008gw,Alabidi:2009bk}). 

In this paper we investigate the effect of the transition
on curvature perturbations on all subhorizon scales, 
including those on sufficiently small scales on which the 
transition cannot be regarded instantaneous any more.
The paper is organized as follows. In Sec.~(\ref{sec:ginv}) we 
derive the governing equation for the evolution of the potential $\Psi$
for an arbitrary equation of state and velocity of sound. 
We review the generation of the perturbations during slow-roll
inflation in Sec.~(\ref{sec:inflation}).
We study the evolution of $\Psi$ through a transition
into radiation and matter domination in 
Secs.~(\ref{sec:rad}) and (\ref{sec:mat}) before concluding.
In Appendices \ref{appA} and \ref{appB}, we present
analytical approximations that we have employed to    
reproduce and understand certain features that we have
found in our numerical estimations.  

Thoughout the paper, we use the Planck units $8\pi G=M_P^{-2}=1$.
\section{Gauge invariant cosmological perturbations}
\label {sec:ginv}

In this section, we introduce the governing equations for the 
scalar perturbations about the homogeneous and isotropic 
background that we will use through out this paper.

The background space-time geometry for a  spatially flat
universe  is given by the FLRW metric,
\be
ds^2=a^2\(\tau\)\(-d\tau^2+dx^2+dy^2+dz^2\)
\,,
\ee
where $\tau$ is the conformal time defined in terms of the
cosmic time $t$ as $d\tau=dt/a$, and $a(\tau)$ is the scale
factor.
Relative to the above coordinates the energy-momentum 
tensor of a perfect fluid at rest is 
\bea
T^0_0&=&-\rho\,,\\
T^i_j&=&P\,\delta^i_j
\,,
\eea 
where $\rho$ and $P$ are the energy density and pressure of 
the fluid.

The governing equations for the system are determined by 
the Einstein Field equations and the energy momentum 
conservation. That is 
\bea
R^{\mu\nu}-\frac{1}{2}g^{\mu\nu}R&=&T^{\mu\nu}
\label{field_eq}\,,\\
T^{\mu\nu}\,_{;\nu}&=&0
\label{conserv_eq}\,.
\eea
In the present case they give the following set of equations:
\bea
&&3\calh^2=\rho a^2\,,
\\
&&\rho'+3\calh\(\rho+P\)=0\,,
\eea
where a prime (${~}'$) denotes the differentiation with respect to $\tau$,
and the conformal Hubble parameter $\calh$ is defined by
$\calh=a'/a$, hence is related to the physical Hubble parameter $H$
by $\calh=aH$.
A combination of the above two equations gives a second order
differential equation for $a(\tau)$,
\be
\calh'=-\frac{1}{6}\(\rho+3P\)a^2\,.
\ee

Ignoring the vector and tensor perturbations, the perturbed metric
can be written to first order as
\be
ds^2=a^2\(\tau\)\[-\(1+2A\)d\tau^2-2B_{,i}d\tau dx^i+
\[\(1+2D\)\delta_{ij}+2E_{,ij}\]dx^idx^j\]
\,,
\ee
where a comma denotes the partial differentiation with respect to the 
spatial coordinates, and $A$, $B$, $D$, and $E$ are the functions
that describe the scalar perturbations. 
The scalar part of the perturbations in the components of
the energy momentum tensor for a perfect fluid are
\bea
\delta T^0_0&=&-\delta\rho \,,\\
\delta T^0_i&=&\(\rho+P\)\(B_{,i}+v_{,i}\)\,,\\
\delta T^i_j&=&\delta P\,\delta^i_j\,, 
\eea
where we have decomposed the fluid 3-velocity $v^i$ 
into a divergence free vector and the gradient of a scalar, 
$v^{,i}$, to extract the scalar part of the perturbations as usual. 

It is convenient to express the governing equations in terms of
quantities which are invariant under infinitesimal coordinate 
transformations. Bardeen~\cite{Bardeen:1980kt} gives the 
following two gauge invariant quantities for the metric,
\bea
\Phi&=&A+\frac{1}{a}\[a\(B-E'\)\]'
\,,\\
\Psi&=&D+\calh\(B-E'\)
\,,
\eea
and for the matter variables,
\bea
\delta_C&=&\delta+3\(\frac{\rho+P}{\rho}\)\calh\(v-B\)
\,,\\
\bar v&=&v-E'
\,.
\eea
The functions $\Psi$ and $\Phi$ represent the lapse function
perturbation and the curvature perturbation in the longitudinal gauge 
respectively. For the matter variables $\delta_C$ is the density
contrast $\delta\rho/\rho$ on the comoving slice and $\bar v$ s the 
velocity of matter on the longitudinal gauge relative to the
$x^i=constant$ observers~\cite{Kodama:1985bj}.

For the matter perturbations of scalar type, there is a third gauge
invariant quantity often called the entropy perturbation, which 
can be constructed in terms of the matter variables only:
\be
\delta\bar P=\delta P-\frac{P'}{\rho'}\delta\rho\,.
\ee
The entropy perturbation measures the difference between uniform
density and pressure hypersurfaces. For an equation of state in which
the pressure is a function of the energy density alone
(the so-called barotropic equation of state), 
the uniform density and pressure hypersurfaces coincide and the 
pressure perturbation is called adiabatic.
For a more general equation of state this is no longer the case and 
we get a non-zero entropy perturbation. However, there is one
important exception in which the perturbation is purely adiabatic
although the equation of state is not barotropic. That is the case
of a single real scalar field. In this case, $P=K-V(\phi)$ and 
$\rho=K+V(\phi)$, where $K=\dot\phi^2/2$.
Then, it is known (or easily seen from the equations below) that
one obtains a closed single second order differential equation
for the scalar perturbation if one defines the sound velocity as
the ratio between $\delta P$ and $\delta\rho$ on the comoving 
hypersurfaces (denoted by the suffix $C$),
\begin{eqnarray}
\delta P_C=c_{\rm s}^2\delta\rho_C\,.
\label{vsound}
\end{eqnarray}
In the present case, $c_{\rm s}=1$ since $\delta P_C=\delta\rho_C=\delta K$.
Note that, since $\delta p=c_{\rm s}^2\delta\rho$ in any gauge 
for a barotropic fluid, the definition of the sound velocity
$c_{\rm s}$ by Eq.~(\ref{vsound}) is valid both for a scalar field and
for a barotropic fluid. 

Now we can express the perturbed equations for Eqs.(\ref{field_eq})
and (\ref{conserv_eq}) in terms of the gauge invariant variables 
above as
\bea
\nabla^2\Psi&=&-\frac{\rho\,a^2}{2}\,\delta_C
\,,\\
\Phi+\Psi&=&0
\,,\\
\(\Psi a\)'&=&-\frac{1}{2}\(\rho+P\)a^2(a\bar v)
\,,\\
\frac{1}{a}\(a\bar v\)'&=&-\Psi+\frac{1}{\(\rho+P\)}\,\delta P_C
\,.
\eea
Then setting $\delta P_C=c_{\rm s}^2\rho\,\delta_C$ in the last
equation, we can reduce the equations above to the
following second order differential equation for $\Psi$,
\be\label{Psi_evol1}
a^2\(\rho+P\)\[\frac{f'_{\bm k}}{a^2\(\rho+P\)}\]'
+\left[k^2c^2_{\rm s}-\frac{1}{2}\left(\rho+P\right)a^2\right]f_{\bm k}=0\,,
\ee
where $f_{\bm k}=a\Psi_{\bm k}$.
The equation above determines the evolution of the potential $\Psi$
for a given equation of state and velocity of sound $c_{\rm s}(\tau)$.
In the following sections we use Eq.~(\ref{Psi_evol1}) to 
calculate the evolution of the perturbations generated during slow-roll
inflation through a rapid transition into radiation domination
and matter domination epochs.
%

\section{The generation of perturbations during inflation}
\label{sec:inflation}

We assume the universe is dominated by a minimally coupled
real scalar field $\phi$ during inflation.
As mentioned in the previous section, for this universe,
the energy density and pressure of the
homogeneous scalar field $\phi\equiv\phi(\tau)$ are given by
 \bea
 \rho=K+V\(\phi\)\,,\quad P=K-V\(\phi\)\,;
\quad K\equiv\frac{1}{2}\dot\phi^2=\frac{1}{2a^2}\phi'^2 \,. 
 \eea
The density and pressure perturbations are then related by
\be
\delta P=\delta\rho-2V_{,\phi}\delta\phi\,,
\ee
which leads to $\delta P_C=\delta\rho_C$ on comoving hypersurfaces
and hence $c_{\rm s}=1$.
During this stage, the evolution equation for $f_{\bm k}$,
Eq.~(\ref{Psi_evol1}), may be written as
\be\label{fk_evol}
f''_{\bm k}-2\frac{\phi''}{\phi'}f_{\bm k}'+\[k^2+\calh'-\calh^2\]f_{\bm k}=0
\,.
\ee

Introducing a new variable $u$ given by
%
\be\label{u_var}
u_{\bm k}=f_{\bm k}\,{\rm exp}\(-\int^\tau\frac{\phi''}{\phi'}d\tau\)
=2kf_{\bm k}\phi'^{-1}=\frac{2k}{\dot\phi}\Psi_{\bm k} \,,
\ee
the differential equation for $f$ can be expressed in the form,
\be\label{wave_eq}
u_{\bm k}''+W^2\(k,\tau\)u_{\bm k}=0,
\ee
where $W^2\(k,\tau\)$ is given by
\be\label{q_func}
W^2\(k,\tau\)=k^2c_s^2+m_{eff}^2a^2\,;
\quad m_{eff}^2a^2=\calh'-\calh^2
-\(\frac{\phi''}{\phi'}\)^2+\(\frac{\phi''}{\phi'}\)'\,.
\ee
The coefficient $2k$ in the definition of $u_{\bm k}$ is chosen
so that the function $a\,q_{\bm k}$ becomes a properly normalized mode
function for an effective scalar field when quantized~\cite{Sasaki:1986hm}.
To be more specific, the field $Q\equiv a q$ becomes a
conformally coupled scalar field with its effective mass-squared
$m_{eff}^2$ defined in the above equation.

We can find approximate analytical solutions for $u_{\bm k}$ 
as follows. We first introduce the following set of slow-roll parameters
\bea
\epsilon_H&=&\frac{\dot\phi^2}{2H^{2}}
=-\frac{a}{\calh^2}\left(\frac{\calh}{a}\right)'
\label{epsroll}\,,\\
\eta_H&=&-\frac{\ddot\phi}{H\dot\phi}
=-\frac{a}{\calh\phi'}\left(\frac{\phi'}{a}\right)'
\label{etaroll}\,.
\eea
Then we rewrite the function $W^2(k,\tau)$ as 
\be
W^2\(k,\tau\)=k^2-
\calh^2\(2\epsilon_H-\eta_H+\eta_H^2-\epsilon_H\eta_H+\frac{\eta'_H}{\calh}\)
\,.
\ee
It is well know that the differential equations for the evolution of 
cosmological perturbations have analytical solution if 
$\epsilon_H$ and $\eta_H$ are constant
\cite{Lyth:1991bc,Stewart:1993bc}. 
Typically the slow-roll parameters are small, at least for the 
range of field values at which perturbations on scales constrained
by observation are generated. 
Their time derivatives then, which are given by
\bea
\frac{1}{H}\dot\epsilon_H&=&2\epsilon_H\(\epsilon_H-\eta_H\)
\,,\\
\frac{1}{H}\dot\eta_H&=&\eta_H\(\frac{\dddot\phi}{H\ddot\phi}+\epsilon_H+\eta_H\)
\,,
\eea
are small, and $\epsilon_H$ and $\eta_H$ can be approximately 
taken as constant, as least for a certain number of $e$-folds.
In that case the conformal time is given by
\be\label{conf_time_inf}
\tau=\int\frac{da}{a^2H}=
-\frac{1}{aH}+\int\frac{\epsilon_Hda}{a^2H}=
-\frac{1}{aH}\frac{1}{1-\epsilon_H}
\ee
where for the time being we have set the integrating constant 
to zero. Writing down explicitly the time dependence in 
Eq.~(\ref{wave_eq}), it becomes a differential Bessel equation, 
\be\label{Bessel_eq}
u''_{\bm k}+\[k^2
-\frac{\(\nu^2-\frac{1}{4}\)}{\tau^2}\]u_{\bm k}=0\,,
\ee
where the order $\nu$ is approximately given by
\be\label{nu_s-roll}
\nu\simeq\frac{1}{2}+2\epsilon_H-\eta_H
\,.
\ee 

In the limit $k\gg aH$, we assume that the fluctuations of the field
are in the Minkowski vacuum,
\be
u_{\bm k}\rightarrow\frac{1}{\(2k\)^{1/2}}e^{-ik\tau}\,.
\ee
Then the solution for $u_{\bm k}$ can be expressed in terms of
the Hankel function of the first kind $H^{(1)}_\nu(x)$, and apart
from an irrelevant  constant phase factor,
$\Psi_{\bm k}$ can be written as
\be\label{Psi_sol_inf}
\Psi_{\bm k}=\frac{\sqrt{\pi}}{2k^{3/2}}\,
\frac{H^2}{\dot\phi}\epsilon_H\(-k\tau\)^{1/2}H^{\(1\)}_\nu\(-k\tau\)
\,.
\ee

On scales well outside the horizon, $k\ll aH$, the amplitude
of $\Psi$  takes the following asymptotic limit,
\be
\Psi_{\bm k}\(\tau\)\simeq\frac{2^{\nu-1}}{\sqrt{2k^3}}
\frac{\Gamma\(\nu\)}{\Gamma\(1/2\)}
\epsilon_H\(-k\tau\)^{1/2-\nu}
\frac{H^2}{\dot\phi}
\,. 
\ee    
Evaluating the expression above at a given time, for example at
the end of inflation, it can be seen that the scale dependence on
superhorizon scales is given by the deviation of  $\nu$ from the
value $1/2$. 
Nevertheless, the scale dependence then is commonly given
by the small variation of the amplitude,
$(H^2/\dot\phi)\vert_{k=aH}$, when perturbations on
different scales leave the horizon during inflation.
In this case one can write
\be
\Psi_{\bm k}\(\tau\)\simeq\frac{2^{\nu-1}}{\sqrt{2k^3}}
\frac{\Gamma\(\nu\)}{\Gamma\(1/2\)}
\epsilon_H\(1-\epsilon_H\)^{1/2-\nu}\,
\(\frac{H^2}{\dot\phi}\)_{k=aH}
\,. 
\ee    

For scales that remain inside the horizon during the
inflationary epoch, $\Psi$ does not display significant
scale dependence even if the slow-roll parameters are
relatively large. For $k\tau>1$, $\Psi_{\rm}$ converges
very quickly to
\be
\Psi_{\bm k}\(\tau\)=\frac{1}{\sqrt{2k^3}}
\frac{H^2}{\dot\phi}\epsilon_H
\,{\rm e}^{-ik\tau}
\,,
\ee
and only perturbations on scales for which $k\tau\simeq1$,
may vary respect to the scale invariant case.

A special case, in which the slow-roll parameters defined above are
constant and we get an exact analytical solution for $\Psi_{\bm k}$, 
is power law inflation. In power law inflation $a\propto t^q$, and 
the slow-roll parameters give
\be
\epsilon_H=\eta_H=\frac{1}{q}\,.
\ee
Then the solution for $\Psi$ is given by Eq.~(\ref{Psi_sol_inf}) with
\be
\nu=\frac{1}{2}\frac{\(q+1\)}{\(q-1\)}
\,.
\ee 
For large $q$, when the slow-roll
conditions hold strongly, $\Psi_{\bm k}$ on superhorizon scales
becomes
\footnote{The relation in Eq.~(\ref{lim})  holds barring an exact de
Sitter background, which corresponds to $\dot\phi=0$. 
Note that we are considering the quantity
$(H^2/\dot\phi)$ small to be consistent with observation.}
\be\label{lim}
\Psi_{\bm k}\(\tau\)=\frac{1}{\sqrt{2k^3}}
\frac{1}{q}\,
\(\frac{H^2}{\dot\phi}\)_{k=aH}
\simeq0
\,.
\ee
%

On cosmological scales, which leave the horizon long 
before the end of inflation, it is common to describe the small 
inhomogeneties observed in CMB temperature fluctuations by
the curvature perturbation on comoving hypersurfaces, 
$\calr$. The curvature perturbation $\calr$ is related to
the Bardeen potential $\Psi$ by 
\be\label{curv_p}
\calr=
-\frac{2}{3}\frac{\calh^{-1}\Psi'+\Psi}{\(1+w\)}-\Psi
\,,
\ee  
where $w=P/\rho$ and $(1+w)=2\epsilon_H/3$ during inflation.

For power law inflation, the derivative of $\Psi$ can be expressed
as
\be
\calh^{-1}\Psi_{\bm k}'\(\tau\)
=\frac{\sqrt{\pi}}{2k^{3/2}}
\frac{H^{2}}{\dot\phi}\epsilon_H\(-k\tau\)^{1/2}
\[\(-k\tau\)\(1-\epsilon_H\)H^{\(1\)}_{\nu+1}\(-k\tau\)-
\(1+\epsilon_H\)H^{\(1\)}_{\nu}\(-k\tau\)\]
\,.
\ee
Then we find the following exact expression for $\calr$:
\be\label{curv_sol_inf}
\calr_{\bm k}\(\tau\)=-\frac{\sqrt{\pi}}{2k^{3/2}}\,
\frac{H^{2}}{\dot\phi}\(1-\epsilon_H\)\(-k\tau\)^{3/2}
H^{\(1\)}_{\nu+1}\(-k\tau\)
\,.
\ee

On superhorizon scales,   taking the asymptotic
limit of the Hankel function, $(-k\tau)\rightarrow0$, 
the curvature perturbation can be expressed as
\be\label{r_plaw}
\calr_{\bm k}=\frac{2^{\nu-1/2}}{\sqrt{2k^{3}}}
\, \frac{\Gamma\(\nu+1\)}{\Gamma\(3/2\)}
\(\nu+1/2\)^{-\nu-1/2}
\(\frac{H^2}{\dot\phi}\)_{k=aH}
\,,
\ee
in agreement with \cite{Lyth:1991bc}.
In the limit $q\rightarrow\infty$, we get 
\be
\calr_{\bm k}=\frac{1}{\sqrt{2k^3}}\(\frac{H^2}{\dot\phi}\)_{k=aH}
\,,
\ee
which is the standard result for a slowly varying
potential~\cite{Sasaki:1986hm}. For $\epsilon_H$ not too small this result
is modified by the factor in Eq.~(\ref{r_plaw}).

In view of the expressions derived above for 
$\calr$ and $\Psi$, it is clear that for small values
of the slow-roll parameters the amplitude of $\Psi$
is much smaller than $\calr$ on superhorizon scales.
On scales well inside the horizon, $k>aH$,
the curvature perturbation $\calr$ increases with the
comoving scale as $k/aH$, while $\Psi$ is scale independent
so the amplitude of $\Psi$ is further suppressed
relative to $\calr$.  
It is natural then to 
neglect $\Psi$ respect to $\calr$ during inflation if the
slow-roll parameters are not too close to 1.

As we stressed above, Eqs.~(\ref{Psi_sol_inf}) and (\ref{curv_sol_inf})
are exact solutions for the perturbed quantities $\calr$ and $\Psi$
 generated during power law inflation.
They provide approximate solutions for slow-roll inflation
if we consider a suitable power law for the scale factor of the
homogeneous background during a sufficiently short lapse of time.
The approximation will be valid then for certain number
of $e$-folds, depending on the variation of the slow-roll
parameters.
Here we are interested in the behaviour of the
perturbations at the end of inflation, and therefore
the power law solutions at that time may differ considerably 
from those at the time of horizon crossing for scales which 
are well outside the horizon at the end of inflation.
Nevertheless, since we focus on the scales smaller than the
horizon, this problem does not affect our analysis.

In the following sections, we give approximate analytical
expressions for $\Psi$ for a transition from inflation to
radiation and matter domination alongside of numerical
calculations. For the numerical estimation it is convenient
to normalize the solutions of $\Psi$ with respect to the value of
$\calr$ at horizon crossing at the end of inflation. Thus we
define this value, $\cala_\calr$, as
\be\label{Rnorm}
\cala_\calr\equiv\sqrt{2k^3}\calr_{k=\calh_{\rm e}}=
r\(q\)\(\frac{H^2}{\dot\phi}\)_{k=\calh_{\rm e}}
\,,
\ee
where $r(q)$ is just the factor in Eq.~(\ref{r_plaw}).
That is
\be
r\(q\)=2^{\nu-1/2}\, \frac{\Gamma\(\nu+1\)}{\Gamma\(3/2\)}
\(\nu+1/2\)^{-\nu-1/2}
\,.
\ee
  
\section{The transition into radiation domination}
\label{sec:rad}

In this section, we study the behavior of the potential $\Psi$
during a sharp transition into radiation domination from a 
slow-roll inflationary epoch.
The evolution of the potential $\Psi$ from 
Eq.~(\ref{Psi_evol1}) is determined by 
\be\label{Psi_evol2}
f''_{\bm k}-\left[\frac{\left(\rho+P\right)'}{\left(\rho+P\right)}
+2\calh\right]f'_{\bm k}
+\left[k^2c^2_{\rm s}\left(\tau\right)
-\frac{1}{2}\left(\rho+P\right)a^2\right]f_{\bm k}=0
\,.
\ee

It is necessary to do a constant time shift of the conformal
time to match the scale factor between inflation and the 
radiation dominated stage. If the transition into radiation
domination occurs much more rapidly than a Hubble time,
then the scale factor at the end of inflation is approximately
given by  $a_{\rm e}\simeq1/H_{\rm e}\tau_{\rm e}$, 
and we have to replace in the formulae for the perturbations
during inflation $\tau$ by $\tau-\tau_R$, where $\tau_R$ is the constant
conformal time shift. Neglecting $\epsilon_H$ in
Eq.~(\ref{conf_time_inf}),  we get $\tau_R=2\tau_{\rm e}$.

One can in principle solve Eq.~(\ref{Psi_evol2}) numerically for
a given equation of state and velocity of sound.
Here we are interested in the qualitative behaviour of
$\Psi$ for a sharp transition into radiation domination,
so we assume that the equation of state changes abruptly to reach
its radiation domination value in some small fraction of a Hubble time.
We can parametrise then the equation of state as $P=w(t)\rho$, where
$w(t)$ is a step like function with a certain width, 
$\Delta t=\delta/H_{\rm e}$.
During inflation $(1+w)=2\epsilon_H/3$, 
which is small quantity under slow-roll conditions. For a rapid transition
into radiation domination it increases very sharply until it reaches the
value $4/3$, corresponding to the radiation equation of state
$P=\rho/3$.

To solve Eq.~(\ref{Psi_evol2}) for $f=a\Psi$, it is convenient
again to express it in terms of $u$, 
$u_{\bm k}=2(k/a)(\rho+P)^{-1/2}f_{\bm k}$.
The evolution equation for $u$ takes the form,
\be\label{eq_transr}
u_{\bm k}''+W^2\(k,\tau\)u_{\bm k}=0
\,,
\ee  
where $W^2(k,\tau)$ is
\be\label{q_eq}
W^2\(k,\tau\)=k^2c^2_s\(\tau\)-4\pi G\left(\rho+p\right)a^2-
\frac{1}{4}A^2\(\tau\)+\frac{1}{2}A'\(\tau\)
\,,
\ee
and $A(\tau)$ is defined by
\be
A\(\tau\)=\left[\frac{\left(\rho+p\right)'}
{\left(\rho+p\right)}+2\calh\right]\,.
\ee

\begin{figure}
\begin{center}
\includegraphics[angle=0,width=0.95\textwidth]{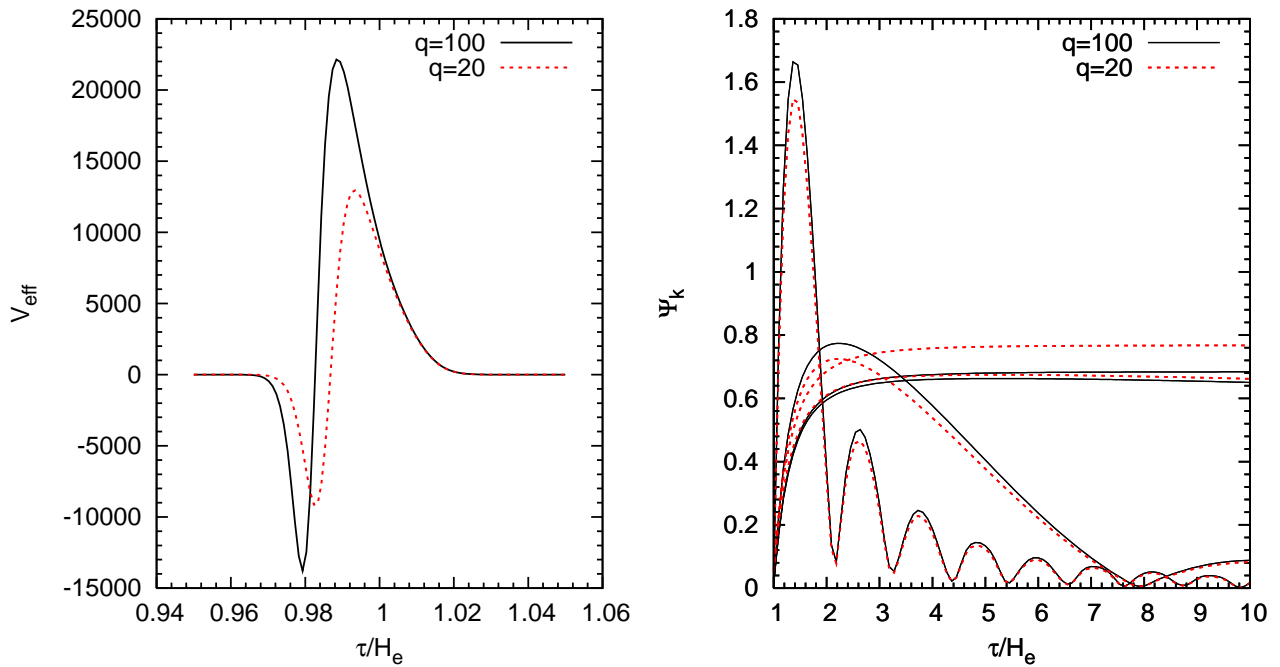}
\end{center}
\caption{\label{Fig:1}
\scriptsize The figure on the left shows $V_{\rm eff}$
through a transition into radiation domination as a 
function of $\tau$ in $H_{\rm e}^{-1}$ units for two
different values of $q$ in power law inflation.
The transition time corresponds to $\delta=0.01$.
The figure on the right shows the evolution of  
 $\sqrt{2k^3}\vert\Psi_{\bm k}\vert$ in $\cala_\calr$
units for 4 different modes with 
$\alpha\equiv k/a_eH_e=5, 1, 0.1, 0.01$, and
the higher frequency modes corresponding to larger wavenumbers $k$.
The parameters are the same as the left figure.}
\end{figure}

During the transition we may write $W^2(k,\tau)$ as
\be\label{freq_trans}
W^2\(k,\tau\)\simeq k^2c^2_{\rm s}\(\tau\)
-V_{\rm eff}\(\tau\)
\,,
\ee
where $V_{\rm eff}$ is approximately given by
\be\label{Veffective}
V_{\rm eff}\(\tau\)\simeq-\frac{3}{4}\(\frac{w'}{1+w}\)^2
+\frac{1}{2}\frac{w''}{\(1+w\)^2}
-\calh\frac{w'}{1+w}
\,,
\ee
neglecting all the terms proportional to $\calh^2$.
The terms proportional to the derivatives of $w$ are sharply
peaked functions, with an approximate amplitude of order
$\calh^2/\delta^2$ during the transition when the 
equation of state varies.
In  Fig.~\ref{Fig:1}, we show the time dependence of 
$V_{\rm eff}(\tau)$ for an equation of state such that during
the transition $(1+w)$ it behaves like
\be
\(1+w\)=\frac{2}{3}\[1+{\rm erf}
\(\frac{\calh_{\rm e}\(\tau-\tau_{\rm e}\)}{\delta}\)\]
\ee
where we have ignored the small value proportional to $\epsilon_H$
it has during inflation. 
We plot $V_{\rm eff}$ for $\delta=0.01$ and two different values
of $q$.  The height and width of the peaks shown in the figure 
depend on $\delta$ as well as $\epsilon_H$, which is evident 
from the expression of $V_{\rm eff}$ in Eq.~(\ref{Veffective}).
As we shall see, the variation of $V_{\rm eff}$ with $\epsilon_H$
is responsible for the peculiar scale dependence of the spectrum
of $\Psi_{\bm k}$ on subhorizon scales. On superhorizon scales,
one gets the usual red tilt on the spectrum for power law inflation.

Once the equation of state reaches its radiation domination
constant value,  the derivatives of $w$ vanish and 
$W^2(k,\tau)$ becomes
\be
W^2\(k,\tau\)=\(k^2c^2_{\rm s}-\frac{2}{\tau^2}\)
\,,
\ee
which gives a Bessel differential equation of order
$\nu=3/2$ for $u$ when $c_{\rm s}$ is constant. Note that
the solutions for half integer values of $\nu$
are just spherical Bessel functions.  

For scales $k\gg \calh/\delta$ then, the gradient term in
$W^2(k,\tau)$ dominates. 
Assuming that  $c^2_{\rm s}(\tau)$ drops down from $1$ to $1/3$
during the transition, an approximate estimate for $\Psi_{\bm k}$
on these small scales, $k\gg aH/\delta$, throughout the transition 
and radiation domination epoch,  is given by the WKB 
solution,
\be\label{Psi_WKB}
\Psi_{\bm k}\(\tau\)\simeq\frac{1}{\(2k\)^{3/2}}
\frac{\(\rho+p\)^{1/2}}{\sqrt{c_{\rm s}\(\tau\)}M_p^2}
{\rm e}^{-ik\int c_{\rm s}\(\tau\)d\tau}
\,, 
\ee
where we have recovered the Plank mass, $M_p=(8\pi G)^{-1/2}$,
so that the dimensionless nature of $\sqrt{2k^3}|\Psi_k|$ is explicit.
Hereafter, we call this non-dimensional amplitude as the amplitude
of $\Psi$ at a comoving scale $k$.
The solution in this regime represents plane waves
with an increasing amplitude during the transition that
starts to decay once the radiation conditions ensue.
The maximum amplitude of the evanescent wave
depends only on the energy scale of inflation, 
$\Psi_M\sim H_{\rm e}/M_p$.

In the opposite regime, $k\ll \calh_{\rm e}/\delta$, the variation
of the equation of state from the slow-roll regime is responsible 
of the growth of the perturbed quantity $\Psi$.
We can neglect the terms proportional to $f_{\bm k}$ in
Eq.~(\ref{Psi_evol1}), and therefore it becomes
\be
f''_{\bm k}-\[\frac{\(\rho+p\)'}{\(\rho+p\)}+2\calh\]f'_{\bm k}=0
\,,
\ee  
which gives the following conserved quantity,
\be\label{int_c}
\frac{f'_{\bm k}}{a^2\(\rho+p\)}={\rm constant}
\,.
\ee
Integrating the equation above, 
we find that $\Psi$ does not change substantially
during the transition if the transition is much faster
than a Hubble time.
In fact, the relative variation of the
potential $\Psi$ during the transition is roughly
\be
\(\frac{\Delta\Psi}{\Psi}\)\sim\frac{\delta}{\epsilon_H}
\,.
\ee
Therefore taking $\Psi$ to be constant during the
transition on these scales appears to be a good
approximation in the limit $\delta\lsim\epsilon_H$.

Neglecting the small variation of $\Psi$ and the scale
factor during the transition, Eq.~(\ref{int_c}) implies that
the curvature perturbation on comoving hypersurfaces, 
$\calr$, remains constant during the transition.
Thus for a rapid transition from slow-roll into radiation domination,
there is a certain range of scales inside horizon for which
the transition may be taken as instantaneous.
Using the junction conditions in 
Eqs.(\ref{junc1}) and (\ref{junc2}),
we get the result found in \cite{Lyth:2005ze},
\be\label{Psi_delta0}
\Psi_{\bm k}\(\tau\)=2\calr_k\(\tau_e\)
\frac{\[\(x-x_e\)\coseno\(x-x_e\)-\(1+xx_e\)\seno\(x-x_e\)\]}{x^3}
\,,
\ee
where $x=c_{\rm s}k\tau$ and 
$x_{\rm e}=c_{\rm s}k\tau_{\rm e}$, with the velocity of sound given
by $c_s=1/\sqrt{3}$.
On superhorizon scales, $x_{\rm e}\to0$ and $x\ll1$, $\Psi$ reaches
the constant value,
\be
\Psi_{\bm k}\simeq-\frac{2}{3}\calr_{\bm k}\(\tau_{\rm e}\)
\,,
\ee
whereas the modes inside the horizon at the end of
inflation undergo damped oscillations.

\begin{figure}
\begin{center}
\includegraphics[angle=0,width=0.95\textwidth]{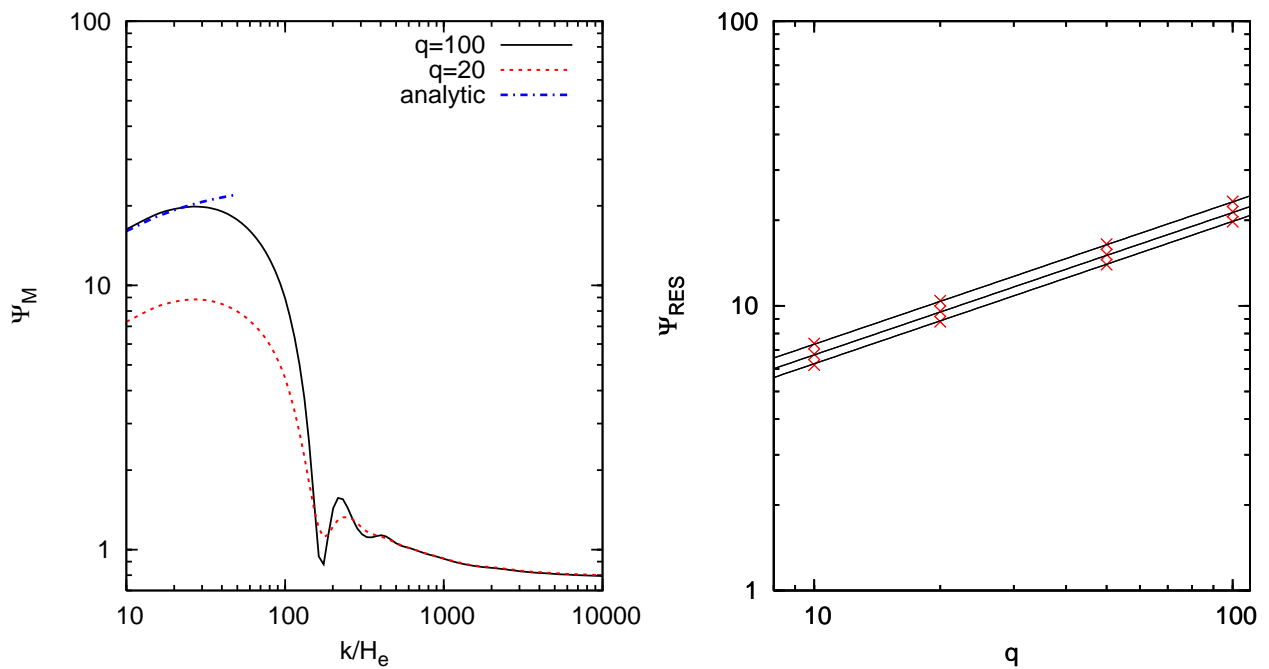}
\end{center}
\caption{\label{Fig:2}
\scriptsize On the left hand side, we show the scale
dependence of the amplitude of $\Psi_M$ in $H_{\rm e}/M_P$
units, defined as $\sqrt{2k^3}\vert\Psi_M(k)\vert$, 
for a transition into radiation domination. 
The two curves are evaluated for $\delta=0.01$.
The comoving wavenumber $k$ is in $H_{\rm e}^{-1}$
units. 
The dashed line represents the maximum of the analytic expression
in Eq.~(\ref{Psi_delta0}).   
On the right pannel, we show the value of the 
maximum for the resonance, $\Psi_{\rm RES}$,
as a function of $q$, for $\delta=0.05\,,0.01$ and $0.001$.
A larger $\delta$ results in a smaller value of $\Psi_{\rm RES}$.
The solid lines represent the interpolation of the numerical
data.
}
\end{figure}

We can check the qualitative behaviour of the 
perturbed potential $\Psi$,
outlined in the preceeding paragraphs,
by solving numerically Eq.~(\ref{Psi_evol2}) for a
definite equation of state which behaves like a 
step function with a given width. Using an error
function for the variation of $(\rho+p)$ during the
transition, with a typical width 
$\Delta\tau\sim\delta/H_{\rm e}$,
we have found that the numerical solutions agree very
well with our analytical expectations in both regimes.
In Fig.~\ref{Fig:1}, we show the evolution of $\Psi$ 
for two different values of $q$ in power law inflation 
with comoving wavenumber $k\ll\calh_{\rm e}/\delta$.
We take values of $\delta$ such that $\delta\lesssim1/q$.
We plot the modes with $\alpha=0.01$, $0.1$, $1$ and $5$, where 
$\alpha=k/a_{\rm e}H_{\rm e}=k/\calh_{\rm e}$.
Thus the former two correspond to superhorizon perturbations
at the end of inflation, while the latter two to subhorizon perturbations.
The superhorizon modes reach the value 
$\Psi_{\bm k}=2\calr_{\bm k}(\tau_{\rm e})/3$ a few Hubble times
after the end of inflation.
For large values of $q$ the scale dependence of these
modes is small. As $q$ decreases 
the deviation from a scale invariant spectrum in the
superhorizon modes becomes more prominent.
On subhorizon scales, $k\gsim\calh_{\rm e}$, $\Psi$
oscillates with decaying amplitude once the radiation
dominated epoch begins.
The relevant quantity to estimate the abundance of
subhorizon PBHs formed at the end of inflation 
is the first maximum of the oscillations,
$\Psi_{\rm M}(k)$ \cite{Lyth:2005ze, Zaballa:2006kh}.
$\Psi_{\rm M}(k)$ increases with $k$, until the comoving scale
approaches $k\sim\calh_{\rm e}/\delta$.
$\Psi_{\rm M}(k)$ then decreases and converges to the
approximate value given in Eq.~(\ref{Psi_WKB}),
$\Psi_{\rm M}\sim H_{\rm e}/M_p$,
as shown in Fig.~\ref{Fig:2}.

The numerical value of $\Psi_{\rm M}(k)$ 
on subhorizon scales for two different values of $q$ is 
shown in the left panel of Fig.~\ref{Fig:2}. 
The value of $\tau$ for the first maximum is approximately
given by 
$\tau_{\rm M}\simeq\tau_{\rm e}+\pi\calh_{\rm e}/2kc_{\rm s}$. 
It is convenient to express  $\Psi_{\rm M}(k)$ in 
$H_{\rm e}/M_p$ units, because in the limit 
$k\gg\calh_{\rm e}/\delta$ it approximately drops 
down to that value.
For a given transition time, $\Delta t\sim\delta/H_{\rm e}$, 
the amplitude of  $\Psi_{\rm M}$ increases as $q$ gets
larger on an interval about the comoving scale
$k\sim\calh_{\rm e}/\delta$, where the resulting spectrum has
a broad resonance.  
As we show in Appendix~\ref{appB}, the resonance can be
interpreted as the effect of  $V_{\rm eff}$ in Eq.~(\ref{freq_trans}) 
acting as a potential barrier for the incoming wave function. 
The higher the potential barrier is the larger is the maximum
amplitude of the resonance.
The width of the resonance is related to the transition time only,
and it is of the order $\calh_{\rm e}/\delta$.
As $k$ becomes much larger than $\calh_{\rm e}/\delta$,
$\psi_{\rm M}$ tends to $H_{\rm e}/M_p$ independently of
the values of $q$ and $\delta$.

In the right panel of Fig.~\ref{Fig:2}, we show numerical values
for  the  peak of the resonance,
which here we denote it as $\Psi_{\rm RES}$,
as a function of $q$ for different values of the transition time. 
The solid lines represent the fitting curve,
\be\label{rad_res}
\sqrt{2k^3}\,\Psi_{\rm RES}\(q,\delta\)=\sqrt{q}\[A+
B\sqrt{\delta}+C\delta\]
\,,
\ee
where $A=2.47$, $B=-4.74$ and $C=-1.58$ in $H_{\rm e}/M_p$
units.
The estimated value of  peak of the resonance above
is in agreement with the result found in \cite{Lyth:2005ze},
where the case of an instantaneous transition into radiation
domination was analyzed. Calculating the maximum of 
Eq.~(\ref{Psi_delta0}), we get \cite{Lyth:2005ze}
\be\label{maxres_an}
\sqrt{2k^3}\,\Psi_{\rm RES}\simeq2\sqrt{3}\,
\(1-\epsilon_H\)
\(\frac{H_{\rm e}^2}{\dot\phi_{\rm e}}\)
\,.
\ee
This is the same value we get in Eq.~(\ref{rad_res}) if we take
the limit $\delta\rightarrow0$ and express $\Psi_{\rm M}$ in terms
of $\cala_\calr$ for $\epsilon_H$ small, noting that during inflation,
\be
\frac{H^2}{\dot\phi}=\frac{1}{\sqrt{2}}\(\frac{H}{M_p}\)
\epsilon_H^{-1/2}
\,.
\ee

We can therefore have a rough idea of the shape of the wide 
resonance if $q$ and $\delta$ are given. The height of the resonance is
of the order of the curvature perturbation at the end of inflation,
$\cala_{\rm\calr}$, while the width is approximately given by $1/\delta$.
It is interesting to note as well that while on superhorizon scales a
smaller value of $q$ tends to enhance the tilt of the spectrum, it 
suppresses the amplitude of the resonance and in general of all the
modes well inside the horizon about the comoving scale 
$k\sim\delta/\calh_{\rm e}$.

\section{The transition into matter domination}
\label{sec:mat}

If an epoch of matter domination succeeds the inflationary 
epoch, then the equation of state is $P\simeq0$ and the 
differential equation for the perturbations during this 
period is
\be\label{matter_dom}
u''_{\bm k}+\(c^2_{\rm s}k^2-\frac{6}{\tau^2}\)u_{\bm k}=0
\,,
\ee 
where the conformal time during the matter domination
period is $\tau=2/\calh$, corresponding to $a\propto\tau^2$,
and the velocity of sound $c_{\rm s}$ is practically zero.
In the limit, $c_{\rm s}k\to0$,
the two independent solutions for $u_{\bm k}$ behave like 
a power law, $u_{\bm k}\propto \tau^3$ and $\tau^{-2}$.
Then the solution for 
$\Psi_{\bm k}$ ($\propto \rho^{1/2}u_{\bm k}$) is given by
\be
\Psi_{\bm k}\(\tau\)\simeq C_{\bm k}+D_{\bm k}\tau^{-5}
\,,
\ee
where the integration constants depend on the comoving
scale $k$. Thus as the universe enters a stage of matter domination,
perturbations on scales for which $c_{\rm s}k\tau$ is small
reach certain constant values a few Hubble times after the end of inflation.

As in the preceding section, we consider that the equation of
state changes very sharply from the inflationary value at the
end of inflation,
determined by the slow-roll parameter $\epsilon_H$ in the case
of power law inflation, to the matter domination value $P=0$. 
The behaviour of the perturbed quantities $\Psi$ and $\calr$
during this transition is similar to the case of the transition into
radiation domination for $k\ll\calh_{\rm e}/\delta$. 
On these scales, we can neglect the gradient terms in 
Eq.~(\ref{Psi_evol1}) and therefore the perturbed quantities
$\Psi$ and $\calr$ do not grow appreciably during the transition if
$\delta\lesssim\epsilon_H$.
However in the opposite regime, $k\gg\calh_{\rm e}/\delta$, 
if we assume that
the velocity of sound $c_{\rm s}(\tau)$ drops down to a negligible 
value during the transition, the potential $\Psi$ oscillates with
a decreasing frequency, given by $W(k,\tau)$ in 
Eq.~(\ref{freq_trans}), until $W(k,\tau)$ reaches zero to eventually
become negative.
At this point, $\Psi$ starts to grow and eventually settle down 
to a constant value during matter domination.  

Assuming that $\Psi$ and $\calr$ do not vary during the 
transition, which is valid for $k\ll\calh_{\rm e}/\delta$
and $\delta\lesssim\epsilon_H$,
and taking the slow-roll limit $\Psi\ll\calr$ during 
inflation, we can determine the integrating constants
$C_{\bm k}$ and $D_{\bm k}$. 
Setting the scale factor at the end of inflation to
$a_{\rm e}\simeq2/\tau_{\rm e}H_{\rm e}$,
the potential $\Psi$ is given by
\be\label{Psi_matter}
\Psi_{\bm k}\(\tau\)\simeq
-\frac{3}{5}\calr_{\bm k}\(\tau_{\rm e}\)
\[1-\(\frac{\tau_{\rm e}}{\tau}\)^{5}\]
\,.
\ee
Thus, a few Hubble times after the end of inflation $\Psi$ reaches the
constant value $\Psi_{\bm k}\simeq3\calr_{\bm k}(\tau_{\rm e})/5$
on these scales.
This is the standard result for superhorizon scales during
matter domination. Here we have shown that 
this result will hold for $k\ll\calh_{\rm e}/\delta$
as long as $\delta\lesssim\epsilon_H$.
For perturbations on scales  $k\gsim\calh_{\rm e}/\delta$,
 $\Psi$ and $\calr$ vary during the transition because 
the wavelength of the perturbations is the order or smaller
than the transition time, 
and therefore the estimates above are no longer valid. 
Nevertheless, $\Psi$ also reaches a constant value during matter
domination which in this case is scale dependent.

Approximate analytical solutions are harder to obtain than in
the case of a transition into radiation domination for 
$k\gg\delta/\calh_{\rm e}$ because the gradient term vanishes
during matter domination.
Neglecting the derivatives of the equation of state,
the evolution equation for $u_{\bm k}$  can be written as
\be\label{eq_transm}
u_{\bm k}''+
\[k^2c^2_s\(\tau\)-\frac{3}{2}\calh^2\(1+w\)\]u_{\bm k}
=0
\,,
\ee  
where both $c^2_s(\tau)$ and $(1+w)$ behave like step like
functions with the same width determined by the parameter
$\delta$.
We have tried 1st order WKB solutions by using Airy functions
to match the solutions between inflation and matter domination.
Despite this solution gives the same scale dependence as our
numerical results,
that is $\Psi_{\bm k}\sim (k/\calh_{\rm e})^{1/2}$,
the amplitude is about a factor of $4$ smaller. 
This is probably due to the fact that the WKB approximation
fails marginally for large $\tau$ if $c_s^2=0$.
An asymptotic matching using higher order WKB solutions would
reduce the error. In Appendix~\ref{appA}, we show 
the details of the first order WKB approximation. 
The solution with the appropriate boundary conditions has 
the form,
\be
\Psi _{\bm k}\(\eta\)\sim
4.4\,\(\tau_*\calh_{\rm e}\)^{-\sqrt{6}}\,
\alpha^{1/2}
\(\frac{H_{\rm e}}{M_p}\)\,,
\ee
where $\alpha=k/\calh_{\rm e}$, and $\tau_*$ is the
turning point of the differential equation. The 
turning point marks the time at which the frequency 
of the differential equation changes sign. In this 
case it is given by $W(k,\tau_*)=0$ with
\be
W^2\(k,\tau\)=k^2c^2_s\(\tau\)-
\frac{3}{2}\calh^2\(1+w\)
\,.
\ee

\begin{figure} 
\includegraphics[angle=0,width=1\textwidth]{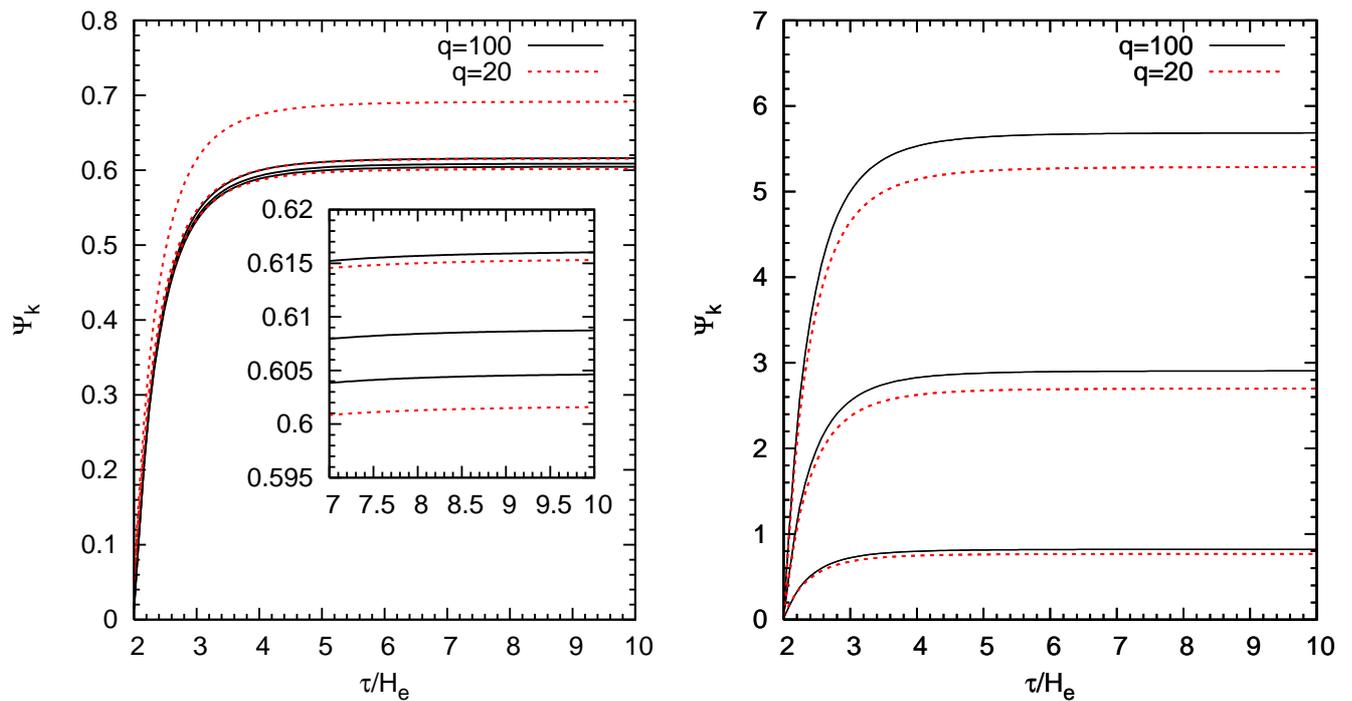} 
\caption{\label{Fig:3} 
\scriptsize The figures show the time evolution of
$\sqrt{2k^3}\vert\Psi_{\bm k}\vert$ in $\cala_\calr$ units through
a transition into matter domination for 3 different modes. On the 
left graph, we show the superhorizon modes 
$\alpha\equiv k/H_{\rm e}=0.2, 0.1$ and $0.01$ for perturbations 
generated during power law inflation.
The smaller wavenumbers result in larger amplitudes as a 
consequence of the red tilt of the spectrum. 
On the right graph, the corresponding modes 
are $\alpha=1, 5$ and $10$. In this case the larger wavenumbers
result in larger amplitudes.
The transition time corresponds to $\delta=0.01$.}
\end{figure}

\begin{figure} 
\includegraphics[angle=0,width=1\textwidth]{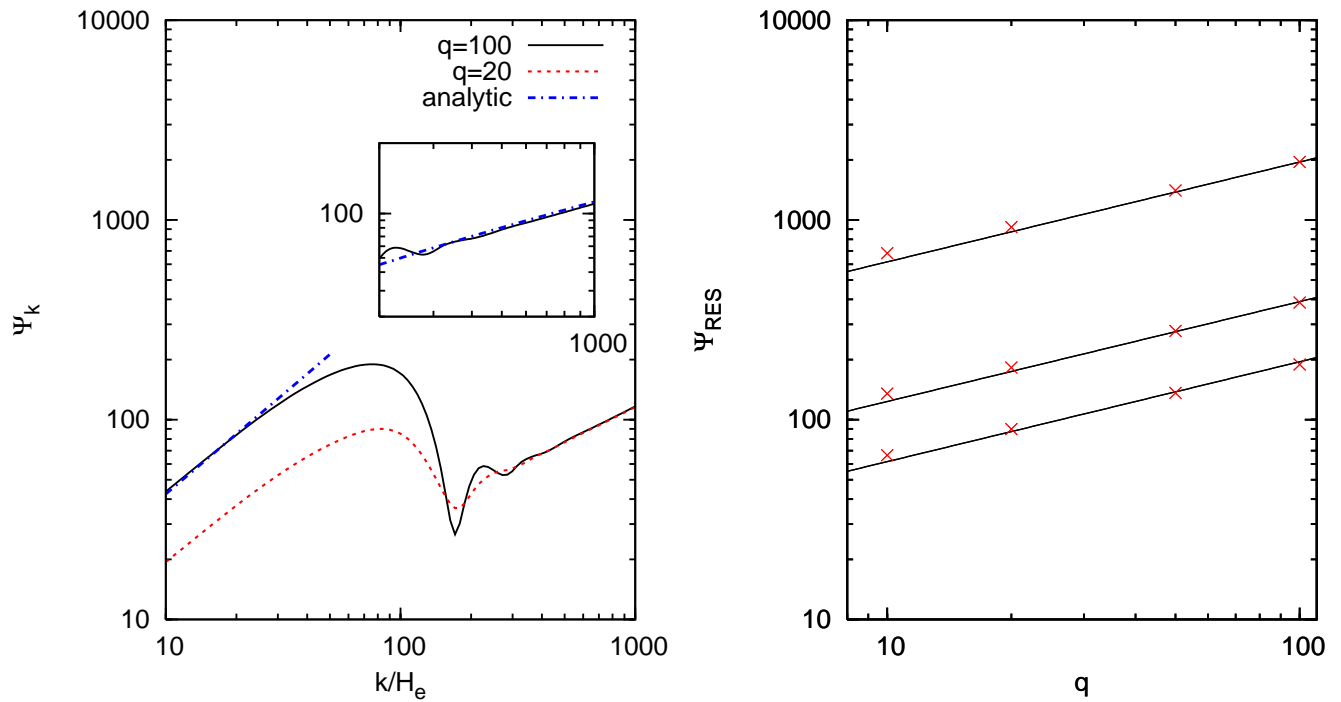} 
\caption{\label{Fig:4}
\scriptsize The figure on the left shows the matter domination 
spectrum of $\Psi$, $\sqrt{2k^3}\vert\Psi_{\bm k}\vert$, 
in $H_{\rm e}/M_p$ units for two different values of $q$.
The transition time is fixed at $\delta=0.01$.
The dashed lines represent our analytical approximations
onto the left and right of the resonance, 
$\Psi_{\bm k}\propto\alpha$ and $\sqrt{\alpha}$,
respectively.
On the right, the peak value at the resonance, $\Psi_{\rm RES}$,
is plotted as a function of $q$ for $\delta=0.05$, 
$0.01$ and $0.001$. The smaller values of $\delta$ result in
larger amplitudes for the resonance.}
\end{figure}

Numerical solutions of the differential equation in 
Eq.~(\ref{Psi_evol1}) are shown in 
Figs.~\ref{Fig:3} and \ref{Fig:4}.
We have modeled the time variation of $P=P(\rho)$ and $c_{\rm s}(\tau)$
as in the previous section by an error function with a transition time interval
of $\Delta\tau\sim\delta/\calh$ until they settle down
to the values corresponding to matter domination.
The numerical results for the perturbed quantity $\Psi$ 
are in good agreement with our expectations based on
the qualitative analysis. In the paragraphs below we
describe our numerical results summarized in
Fig.~\ref{Fig:3} and \ref{Fig:4}.

For large values of $q$, the superhorizon modes do not
show substantial scale dependence and $\Psi$ grows and
converge to an approximate value of $\Psi\sim3\cala_\calr/5$.
As $q$ decreases the scale dependence of these modes
gets larger increasing the red tilt of the spectrum,
as shown in Fig.~\ref{Fig:3}. 
The modes corresponding to scales in the range
$\calh_{\rm e}\lsim k\ll\calh_{\rm e}/\delta$, 
reach the approximate value,
\be\label{Psi_subh}
\sqrt{2k^3}\vert\Psi_{\bm k}\vert\simeq
\frac{3\alpha}{5}\,
\frac{H^2_{\rm e}}{\dot\phi_{\rm e}}=
\frac{3\alpha}{5}\,
q^{1/2}
\(\frac{H_{\rm e}}{M_p}\)
\,,
\ee
taking the asymptotic limit $k\tau_{\rm e}\gg1$ of 
$\calr_{\bm k}$ in Eq.~(\ref{curv_sol_inf}).
On these range of scales $\Psi_{\bm k}$ increases as
the slow-roll parameter $\epsilon_H=1/q$ gets 
smaller. (See the right panels of Fig.~\ref{Fig:3}
and Fig.~\ref{Fig:4}.)

On the opposite regime, $k\gg\calh_{\rm e}/\delta$, 
the amplitude of $\Psi_{\bm k}$ grows with scale as 
$\alpha^{1/2}$.
As $k$ decreases there are a series of small 
oscillations in the spectrum followed by a 
resonance approximately located at 
$k\sim\calh_{\rm e}/\delta$. The resonance is 
as in the radiation domination case the result 
of the effective potential,$V_{\rm eff}$, on the 
modes with wavenumbers about $k\sim\calh_{\rm e}/\delta$. 
The maximum amplitude of the resonance depends 
on the height and width of the effective potential 
$V_{\rm eff}$. These are determined by the parameters 
$q$ (or $\epsilon_H$) and $\delta$. 
For a fixed value of $\delta$, the potential 
barrier described by $V_{\rm eff}$ defined in 
Eq.~(\ref{Veffective}) behaves as 
$V_{\rm eff}\propto1/(1+w)^2$, increasing its 
height as $\epsilon_H$ becomes smaller. 
As the height of the potential barrier increases, 
the effect on the a fixed mode $k$ of the field 
$u$ that propagates across it, is to increase its 
amplitude (as it can be observed in the
left panel of Fig.~\ref{Fig:4}).
We have described in more detail the effect of a
simple squared potential on the amplitude of 
$\Psi_{\bm k}$ in  Appendix~\ref{appB}. Despite its 
simplicity, the model reproduces qualitatively the 
features observed in the spectrum of $\Psi_{\bm k}$
shown in Fig.~\ref{Fig:4}.

Numerical results for the maximum of the resonance in 
the spectrum of $\Psi_{\bm k}$ are shown in the right panel of
Fig~.\ref{Fig:4}. A good fitting curve in the regime shown in the
figure is
\be
\Psi_{\rm RES}\(q,\delta\)=A\,\frac{q^{1/2}}{\delta}
\(\frac{H_{\rm e}}{M_p}\)=
\frac{A\sqrt{2}}{\delta}\,\(\frac{H^2_{\rm e}}{\dot\phi_{\rm e}}\)
\,,
\ee
where $A=0.195$. The dependence of the amplitude of the 
resonance with $q$ is the same as in the radiation domination 
case. 
However, in this case $\Psi_{\rm RES}$ diverges in the limit
of $\delta$ going to zero. This result is not surprising because
if we consider an instantaneous transition and use the junction
conditions~(\ref{junc1}) and (\ref{junc2}), then $\Psi_{\bm k}$
is given by Eq.~(\ref{Psi_subh}) and it would diverge in the 
ultraviolet limit.

In reality, the sound velocity will not be exactly zero
in the matter-dominated stage. 
In the case of a free massive scalar field,
the perturbation will not grow on very small scales
where the wavenumber exceeds the geometrical mean of
the Hubble parameter and the mass, 
$k/a>(k/a)_c\sim\sqrt{Hm}$~\cite{Nambu:1989kh}. 
Furthermore, if there is a $\lambda\phi^4$ 
self-interaction, the critical scale is modified drastically to
$(k/a)_c\sim\lambda^{1/2}m^2/M_P$~\cite{Nambu:1989kh}.
In any case, these scales provide a natural ultraviolet cut-off
and regulate the ultraviolet divergence.

To conclude this section we would like to remark the
similar and different features of the solutions in this case
compared to the transition into radiation domination we studied
in the previous section.
As in the radiation domination case, 
the potential $\Psi$ does not depend
on the slow-roll parameters in the limit $k\gg\calh_{\rm e}/\delta$.
It is only in the opposite regime, when $\calr$ does not vary
considerably during the transition, that its value at horizon crossing
at the end of inflation is relevant for the perturbed quantity 
$\Psi$.
The scale dependence of the solutions for 
$k\lesssim\delta/\calh_{\rm e}$ is similar as well, with large values
of $q$ giving a nearly scale invariant spectrum while increasing
the amplitude of the resonance.
In contrast, although the height of the resonance is proportional to
$\cala_{\calr}$, it depends strongly on $\delta$. 
%


\section{Discussion and Conclusions}

We have studied the behaviour of the primordial perturbations
through a sharp transition from inflation into the radiation and 
matter dominated epochs.
We have found that for transitions that occur much faster than
a Hubble time, there is a range of scales inside the horizon
for which the amplitude of the primordial perturbations 
is enhanced, relative to the amplitude of perturbations that exit
the horizon a few efolds before the end of inflation, if slow-roll
conditions hold during the final stages of inflation.
For a Gaussian distribution of the perturbations, this relatively 
large amplitude increase the probability of strong perturbations
on these small scales, that may  lead to a significant
production of PBHs.

For a transition from inflation into a radiation dominated 
universe, the maximum amplitude of the peculiar gravitational
potential is about
$\Psi\sim2\sqrt{3}H^2_{\rm e}/\dot\phi_{\rm e}$.
Assuming a Gaussian distribution for the primordial 
perturbations, one can estimate the PBH abundance produced at 
the end of inflation, using for example Press-Schechter approach.
(See however \cite{Hidalgo:2007vk} for the effect of 
non-Gaussian perturbations.) 
Given that the observed amplitude of the curvature perturbation
on cosmological scales is roughly given by $\calr\sim10^{-5}$, 
it turns out that unless the spectrum of $\calr$ is 
considerably larger at the end of inflation,  
the amount of PBH production would not have any cosmological
significance \cite{Zaballa:2006kh}.
However, a much larger amplitude on scales $k\sim\calh_{\rm e}$,
is still compatible with current observational data even within
the slow-roll paradigm \cite{Kohri:2007qn} . 
Futhermore, PBHs may be overproduced as well with a more complicated
scale dependence if slow-roll conditions do not hold throughout
the whole period of inflation.

In the case of a transition into a matter dominated universe the 
production of PBHs could be more dramatic even with a very
flat spectrum of $\calr$ on all scales.  
The maximum amplitude for $\Psi$ now depends on delta
as $\Psi\propto 1/\delta$. Therefore if the velocity of sound does
vanish on a very small scale, such that $c_s=0$ for 
$k\sim\calh_{\rm e}/\delta$, 
there may be a significant production of small mass PBHs even if 
the transition time is not very rapid.

Finally we would like to stress that a detailed analysis of specific models,
in which the transition from inflation occurs very rapidly, may reveal the 
overproduction of PBHs, placing constraints on the spectrum of 
the primordial perturbations on scales that are too small for
conventional observations. 

\acknowledgments 

IZ would like to thank P. Chingangbam and K. P. Yogendran for
useful discussions. IZ is grateful as well for the hospitality of
YITP in several occasions where this work started and 
eventually has developed. The work of MS is supported in part
by JSPS Grants-in-Aid for Scientific Research (A)~21244033,
by JSPS Grant-in-Aid for Creative Scientific Research No.~19GS0219,
and by Monbukagaku-sho Grant-in-Aid for the global COE program,
"The Next Generation of Physics, Spun from Universality and Emergence".


\appendix

\section{Matter perturbations on scales 
$k\gg \calh_{\rm e}/\delta$.}
\label{appA}

In this Appendix, we find an approximate analytical
solution for the potential $\Psi$ during the matter domination epoch
on scales such that $k\gg \calh_{\rm e}/\delta$.
As explained in Sec.(\ref{sec:mat}), we assume that the velocity of
sound $c_{\rm s}$ drops down from a value of $1$ during inflation
to zero at the end of inflation much faster than a Hubble time.
In this case the gradient terms vanish and the simple oscillatory
WKB solution we found for the radiation domination case is not valid.
However, we can still find an approximate asymptotic solution
to first order in the WKB expansion that reproduces the scale
dependence of the spectrum of $\Psi$ accurately.

Neglecting the time derivatives in the equation of state,
the evolution equation for  the field $u_{\bm k}$ is given by
\be\label{matt_pert_2}
u''_{\bm k}+\left[k^2c^2_{\rm s}\left(\tau\right)
-\frac{3}{2}\calh^2\left(1+w(\tau)\right)
\right]u_{\bm k}=0
\,,
\ee
where $u_{\bm k}=2k(\rho+p)^{-1/2}\Psi_{\bm k}$.
It is convenient, to introduce an adimensional time parameter
$\eta=\tau\calh_e$, where $\calh_e\equiv\calh(\tau_e)$.
Then the equation of motion may be written as 
\be\label{eq_diff}
\ddot u=Q\(\eta\)u
\,,
\ee
where $Q(\eta)$ is defined as
\be
Q\(\eta\)\equiv
\frac{3}{2}\bar\calh^2\(1+w\)-\alpha^2c^2_s\(\eta\)
\,,
\ee
and $\alpha=k/a_eH_e$ and $\bar\calh=\calh/\calh_e$.

This equation has a turning point $Q\(\eta_*\)=0$ at $\eta_*$. 
For $\eta<\eta_*$,  $Q<0$ and the solutions are oscillatory.
For $\eta>\eta_*$,  $Q>0$ and the two independent solutions
may be given by growing and decaying modes.
Then during inflation and matter domination, the 1st order
WKB solutions are given by
\bea
u^{inf}_\pm\(\eta\)&\sim& \vert Q\(\eta\)\vert^{-1/4}
{\rm exp}\[\pm i\int^{\eta_*}_{\eta}
\sqrt{\vert Q\(\eta\)\vert}
d\eta\]\,,\\
u^{mat}_\pm\(\eta\)&\sim&Q\(\eta\)^{-1/4}
{\rm exp}\[\pm\int^{\eta}_{\eta_*}
\sqrt{Q\(\eta\)}
d\eta\]\,.
\eea

To match asymptotically the solutions above, we can expand 
$Q(\eta)$ to first order about the turning point and use the large
argument expansions for the Airy functions. However, it is simpler
to apply the following Liouville transformation,
\bea\label{Lioville}
Y&=&\(\frac{d\xi}{d\eta}\)^{1/2}u\\
\xi&=&\(\frac{d\eta}{d\xi}\)^2Q\(\eta\)
\eea
where $\xi$ and $\eta$ are analytic functions of each other 
at the turning point. 
Then we get the differential equation for $Y$,
\be\label{Y_eq}
\frac{d^2Y}{d\xi^2}=\[\xi+\vartheta\(\xi\)\]Y
\,,
\ee
where  the function $\vartheta\(\xi\)$ above is 
\be
\vartheta\(\xi\)=\(\frac{d\xi}{d\eta}\)^{-1/2}
\frac{d^2}{d\xi^2}\[\(\frac{d\xi}{d\eta}\)^{1/2}\]
\,.
\ee

The differential equation for $Y$ becomes an Airy differential
equation if we neglect $\vartheta$. An approximate solution for $u$ 
then is given by
\be
\label{match_eq}
u=\(\frac{\xi}{Q}\)^{1/4}\[C_1Ai\(\xi\)+
C_2Bi\(\xi\)\]
\,,
\ee
which we use below to match the first order solutions across the 
turning point.

During inflation $\xi<0$, and then we have
\be
\frac{2}{3}\xi^{3/2}\simeq\int^{-\eta}_{\eta_*} \alpha
c_s\(\eta\)d\(-\eta\)
\simeq k\tau
\,,
\ee
where $\tau\ll\tau_*$ and we have taken 
$c_{\rm s}(\tau)=1$ during inflation.
In the models we consider here the velocity of sound during
inflation always has that value, so from now on we just consider
this case.
In the limit $\eta\ll\eta_*$,  the asymptotic expressions for the 
Airy functions in this limit are 
\bea
Ai\(-z\)&\sim&\pi^{-1/2}z^{-1/4}\seno\(\frac{2}{3}z^{3/2}+\pi/4\)
\,,\\
Bi\(-z\)&\sim&\pi^{-1/2}z^{-1/4}\coseno\(\frac{2}{3}z^{3/2}+\pi/4\)
\,.
\eea

Taking the initial Bunch-Davies vacuum value for the field
 $u_{\bm k}$,
\be
u_{\bm k}\longrightarrow\(\frac{1}{2k}\)^{1/2}e^{-ik\tau}
\,,
\ee
one gets the following values for the integrating constants,
\bea
C_1&\sim&-\(\frac{1}{2k}\)^{1/2}\alpha^{1/2}
\sqrt{\pi}e^{i\pi/4}
\,, \\
C_2&\sim&-C_1
\,.
\eea

On the other hand, during the matter domination period $\xi>0$,
and the asymptotic behaviour of the Airy functions  
in this regime is
\bea
Ai\(z\)&\sim&2^{-1}\pi^{-1/2}z^{-1/4}e^{-\frac{2}{3}z^{3/2}}
\,,\label{Airy_asympt1}\\
Bi\(z\)&\sim&\pi^{-1/2}z^{-1/4}e^{\frac{2}{3}z^{3/2}}
\label{Airy_asympt2}
\,.
\eea

Taking $c_s(\tau)$ and $w(\tau)$ to be zero for $\eta\gg\eta_*$,  
$Q^{-1/4}$  is approximately given by
\be
Q^{-1/4}\(\eta\)
\simeq\frac{1}{\sqrt{2}}\(\frac{2}{3}\)^{1/4}
\eta^{1/2}
\,,
\ee
and therefore the solution for $\Psi$ during the matter domination epoch,
neglecting the decaying mode,  is approximately given by
\be\label{matt_anal_01}
\Psi_{\bm k}=\frac{\(\rho+p\)^{1/2}}{\sqrt{2k^3}M^2_p}
u_{\bm k}\sim
\frac{4.4}{\sqrt{2k^3}}\times\alpha^{1/2}
\eta^{-5/2}{\rm exp}\(\frac{2}{3}\xi^{3/2}\)
\(\frac{H_e}{M_p}\)
\,,
\ee
where we have taken $\tau_{\rm e}=2/\calh_{\rm e}$ and
\be
\frac{\(\rho+p\)^{1/2}}{M^2_p}\simeq
8\sqrt{3}\(\frac{H_{\rm e}}{M_p}\)
\eta^{-3}
\,.
\ee

During the matter dominated epoch, the exponential term in 
Eq.~(\ref{matt_anal_01}) is
\be\label{xi01}
\frac{2}{3}\xi^{3/2}=\int^{\eta}_{\eta*}
\sqrt{\frac{3}{2}\bar\calh^2\left(1+w(\tau)\right)
-\alpha^2 c^2_{\rm s}(\tau)}
\, d\eta
\,.
\ee
By integrating by parts expression Eq.~(\ref{xi01}),
we get
\be\label{xi02}
\frac{2}{3}\xi^{3/2}=\sqrt{6}\[
{\rm ln}\(\eta\)-\int^{\eta}_{\eta_*}
{\rm ln}\(\eta\)g^{-1/2}\frac{dg}{d\eta}
d\eta\]
\,,
\ee
where $g$ is
\be
g\(\eta\)=\[\(1+w\)-\frac{2k^2c^2_{\rm s}}{3\calh^2}\]
\,.
\ee

Neglecting the variation of the logarithmic term in the
integrand , $\xi^{3/2}$ is approximately given by
\be
\frac{2}{3}\xi^{3/2}\sim\sqrt{6}\,
{\rm ln}\(\frac{\eta}{\eta_*}\)
\,,
\ee
and therefore the time variation of $\Psi$ is negligible,
\be\label{matt_anal_02}
\sqrt{2k^3}\Psi_{\bm k}\sim
4.43\times\alpha^{1/2}\(\frac{H_e}{M_p}\)
\eta^{-0.05}\eta_*^{-\sqrt{6}}
\,,
\ee
if $\eta$ is not too large.

This approximate solution is the first order solution in the
WKB expansion. The scale dependence of $\Psi$ on this 
regime is $\alpha^{1/2}=(k/H_{\rm e})^{1/2}$, the same that
we have found on our numerical estimations.
The solution is not time independent, although the time 
dependence is very weak.
On the other hand, the approximate value of the amplitude in 
Eq.~(\ref{matt_anal_02}) results in a smaller value than the 
one we have obtained numerically.
These two differences in the analytic result,  suggest that a 
higher order WKB solution would correct this deficiency. 
In fact, for the 1st order WKB solution 
to be valid uniformly for all $\tau$ it is necessary for $Q(\tau)$ 
to decrease much more rapidly than $1/\tau^2$ when 
$\tau$ goes to infinity \cite{Bender}.
During matter domination  $Q\sim1/\tau^2$, and therefore
we believe that a second or higher order WKB asymptotic 
matching would yield a more accurate result.
Here we are mostly interested in reproducing analytically
the scale dependence of the spectrum,  and when we
compare the analytic with the numeric results in 
Fig.~\ref{Fig:4}, we just regulate the lower limit of the 
integrand in Eq.~(\ref{xi01}) to obtain a reasonable 
estimate. 
We believe that this approach is justified given that a relative 
small error in $\xi$ could in principle result in
a significant error in 
$\Psi\propto {\rm exp}(2\xi^{3/2}/3)$.

\section{The effect of a potential barrier on the potential $\Psi$:
 the broad resonance.}\label{appB}

In this Appendix, we explain qualitatively the origin
of the broad resonance and small oscillations that we have obtained 
in our numerical calculations for the spectrum of $\Psi$.
In the two transitions from inflation studied here, 
the broad resonance is approximately located at the comoving
scale $k\sim \calh_{\rm e}/\delta$ and it is followed by a
series of small oscillations with an amplitude that is
strongly suppressed as $k$ increases
(see the left panels of Figs.\ref{Fig:2} and \ref{Fig:4}).
About those scales, the derivatives of the equation of state
are not negligible, and the function $q^2(k,\tau)$ has a complicated
form that makes difficult to find approximate analytical 
solutions.
For a step like equation of state with a width $\delta/H_{\rm e}$,
these terms typically result  in a sharply peaked function with
an approximate amplitude of $H^2_{\rm e}/\delta^2$.
This suggests that, on scales about $k\sim H_{\rm e}/\delta$, the terms 
proportional to $w'$ in $q^2(k,\tau)$ act as a potential barrier for
the wave equation $u_{\bm k}$ that propagates from the 
inflation to the radiation and matter domination epochs.

In order to reproduce the broad resonance in the spectrum then,
we adopt a square potential acting as a potential barrier, for 
which solutions can be easily found. The model is very crude,
but as we will see it captures the fundamental phenomenon
occurring on such scales. 
We restrict this analysis to the transition from inflation to 
matter domination. This case is easier to analyze because the 
modes of the potential reach a constant value.
The radiation domination modes after the transition from 
inflation oscillate with a decaying amplitude. The quantity of
interest then is the maximum amplitude during the osillations,
which is more cumbersome to calculate. Neverthless, the relative
enhancement of the modes is caused essentially by the same
phenomenon as in the matter domination case.

We are interested in the evolution of the field $u_{\bm k}$  
after the transition into radiation or matter domination
on scales such that $k\gsim H_{\rm e}/\delta$. 
We assume that the velocity of sound, remains constant while
the wave crosses through the potential barrier. 
This is a good approximation for the radiation domination case
because $c_s$ does not vary considerably during the transition.
In the matter domination case, we can consider the idealized 
situation in which $c_s$ drops down to zero once the wave 
function has crossed the potential barrier.
To find an approximation solution for the perturbations during
this epoch  then, we can take the resulting solution as the 
initial condition for (Eq.~\ref{matt_pert_2}). 

With the simplifying assumptions considered above in mind,
we now study the behaviour of the fluctuations in the field 
$u_{\bm k}$ for a square potential. 
The evolution of the field then is determined by 
the differential equation,
\be
u''_{\bm k}+\[\alpha^2-V\(\tau\) \]u_{\bm k}=0
\,,
\ee
where $\alpha\equiv k/H_{\rm e}$, and $V(\tau)$ is given by
\begin{equation}
V=
\left\{
\begin{array}{ll}
0  \,, & \tau<-a\,,\\
\\
v^2 \,, & -a<\tau<0 \,,\\
\\
0 \,, & \tau>0 
\,.
\end{array}
\right.
\end{equation}

Taking the initial value of the field $u_{\bm k}$ corresponding
to the Minkowski vacuum,
\be\label{Minks_vacuum}
u_{\bm k}=\frac{1}{\sqrt{2k}}e^{-ik\tau}
\,,
\ee
the solution for $\tau>0$ is
\be\label{square_sol}
u_{\bm k}=\frac{1}{\sqrt{2k}}
\(Ae^{-ik\tau}+Be^{ik\tau}\)
\,,
\ee
where $q=\sqrt{k^2-v^2}$, and the constants $A$ and
$B$ are given by
\bea
A&=&\frac{\(k+q\)^2-\(k-q\)^2e^{-2iqa}}
{4kqe^{-i\(q-k\)a}}
\nonumber\,,\\
B&=&\frac{v^2\(1-e^{-2iqa}\)}
{4kqe^{-i\(q-k\)a}}
\,.
\eea

As in the previous section, an approximate asymptotic
solution valid for throughout the transition and the matter
domation epoch is
\be\label{match_eq2}
u=\(\frac{\xi}{Q}\)^{1/4}\[D_1Ai\(\xi\)+
D_2Bi\(\xi\)\]
\,,
\ee
where $D_1$ and $D_2$ are integrating constants.
Taking the asymptotic expressions of the Airy functions for
$\tau\ll\tau_*$ in Eqs.~(\ref{Airy_asympt1}) and 
(\ref{Airy_asympt2}), we find that the initial conditions
in Eq.~(\ref{square_sol}) are satisfied if
\bea
D_1&\sim&\frac{1}{\sqrt{2k}}\pi^{1/2}\alpha^{1/2}
\(B-A\)
\,,\\
D_2&\sim&\frac{1}{\sqrt{2k}}\pi^{1/2}\alpha^{1/2}
\(A+B\)
\,.
\eea
The asymptotic solution for $\tau\gg0$ during matter
domination for the potential $\Psi_{\bm k}$ is 
approximately given by
\be\label{manalytic_fit}
\Psi_{\bm k}=\frac{\(\rho+p\)^{1/2}}{2k}
u_{\bm k}
\sim\frac{1}{\sqrt{2k^3}}
4.4\eta_*^{-\sqrt{6}}\,
\vert F\vert\,\alpha^{1/2}
\(\frac{H_{\rm e}}{M_p}\)
\,,
\ee 
where $\alpha=k/H_{\rm e}$, $\eta_*=\tau_* H_{\rm e}$ 
denotes the time at the turning point 
(see Appendix~\ref{appA}), and $F\equiv A+B$.
The function $\vert F\vert$ which modulates the 
solution $\Psi\propto\alpha^{1/2}$ we have found in 
Appendix~\ref{appA} can be written as
\be
\vert F\vert^2=\frac{\alpha^2-v^2{\rm cos}^2
\(l\sqrt{\alpha^2-v^2}\)}{\alpha^2-v^2}
\,,
\ee
for $\alpha>v$, which is the region of the 
comoving scale that we are intersted in reproducing.

In the limit $k\gg H_{\rm e}$ we get the trivial
result $\vert F\vert\simeq1$, which corresponds to 
$A\simeq1$ and $B\simeq0$.
On this regime, the variation in the potential does 
not affect the propagation of the field $u_{\bm k}$,
and $\Psi_{\bm k}\propto\alpha^{1/2}$.
As the comoving scale $k$ decreases $\Psi_{\bm k}$ 
decreases until $k$ approaches the scale close to 
the height of the potential $\alpha=v$. 
About that scale, the amplitude  of $\Psi_{\bm k}$ 
increases achieving its maximum amplitude at scale
about $\alpha\sim v$. For $\alpha\lesssim v$, 
the amplitude of the perturbations decrease again.
The overall effect then is that we find a broad
resonance on the spectrum of $\Psi_{\bm k}$ 
approximately located at the scale $\alpha\sim v$.

\begin{figure}
\begin{center}
\includegraphics[angle=0,width=0.65\textwidth]{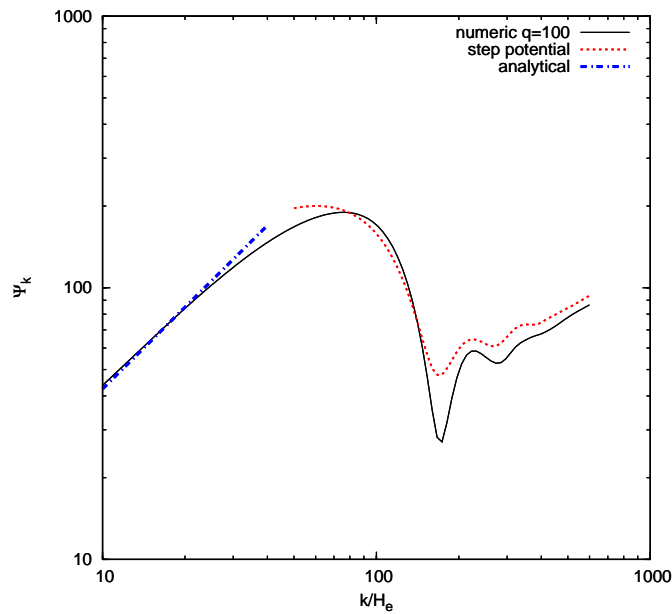}
\end{center}
\caption{\label{Fig:5}
\scriptsize The figure on the left we compare the
matter domination spectrum of  $\Psi_{\bm k}$ we have 
obtained numerically with the anlytical approximation
in Appendices \ref{appA} and \ref{appB}.  
The numerical solution corresponds to $\delta=0.01$
and $q=100$. On the right of the resonance, 
we have found a reasonable fit with
the anlytical approximation for $v=118$, 
$l=0.025$ and $\eta_*=0.2$.
The dashed line represents our analytical approximation
onto the left of the resonance, $\Psi\propto\alpha$.
}
\end{figure}

In Fig.~\ref{Fig:5}, we compare the result of our
numerical calculation of the spetrum of 
$\Psi_{\bm k}$ with the analytical approximation
above in Eq.~(\ref{manalytic_fit}).  
The numerical result is for $\delta=0.01$ and 
$q=100$. To the left of the resonace, we have
plotted the analytical approximation for 
$k\lesssim H_{\rm e}$ in Eq.~(\ref{Psi_subh})
which is a good approximation on that scales.
To the right of the resonance, on scales such 
that $k\gsim H_{\rm e}/\delta$, we show the 
analytical approximation above in 
Eq.~(\ref{manalytic_fit}) for $v=118$ and 
$\delta=0.02$. 
To regulate the integral in 
Eqs.(\ref{xi01}) and (\ref{xi02}) we have
used the lower limit of integration 
$\eta_*=0.2$, which gives the right result
for the scales shown in Fig.~\ref{Fig:5}. 
For these values of the parameters $v$ and
$\delta$, we see that the square potential
reproduces the resonance we have found in our 
numerical calculations sufficiently accurately
given the crude analytical approximation we 
have done.

\end{document}